\documentclass[aip,showpacs,preprint,rsi]{revtex4-1}
\usepackage{graphicx}
\usepackage{latexsym}
\usepackage{amssymb}
\usepackage{amsmath}
\usepackage{color}

\begin{document}
\title{Efficient Radio Frequency filters for space constrained cryogenic set-ups}

\author{Soumen Mandal}
\email{soumen.mandal@gmail.com}
\author{Tobias Bautze}
\author{R\'emi Blinder}
\author{Tristan Meunier}
\affiliation{Institut N\'eel, CNRS and Universit\'{e} Joseph Fourier, 38042 Grenoble,
France}
\author{Laurent Saminadayar}\email{saminadayar@grenoble.cnrs.fr}
\affiliation{Institut N\'eel, CNRS and Universit\'{e} Joseph Fourier, 38042 Grenoble,
France} \affiliation{Institut Universitaire de
France, 103 boulevard Saint-Michel, 75005 Paris, France}
\author{Christopher B\"{a}uerle}
\email{bauerle@grenoble.cnrs.fr} \affiliation{Institut N\'eel, CNRS and Universit\'{e} Joseph Fourier, 38042 Grenoble,
France}


\begin{abstract}
\section*{Abstract}
Noise filtering is an essential part for measurement of quantum
phenomena at extremely low temperatures. Here, we present the design
of a filter which can be installed in space constrained cryogenic
environment containing a large number of signal carrying lines. Our
filters have a -3db point of 65kHz and its performance at GHz
frequencies are comparable to the best available RF filters.
\end{abstract}

\pacs{84.30.Vn}

\maketitle
\section{Introduction}
With the advancement of science, low noise measurements at low
temperatures are becoming increasingly important. Experiments
probing quantum mechanical phenomena require high degree of immunity
from stray noise present in our surroundings. Though it is
relatively easy to reach temperatures as low as 10mK in commercial
dilution refrigerators, achieving similar electronic temperatures on
the other hand is far from being trivial. This is due to the fact
that the microscopic system to be studied is connected to
measurement instruments at room temperature. The electromagnetic
noise coming from the room temperature environment perturbs the
system through propagation through the measurement lines. To
eliminate such disturbance, effective cryogenic filters are
necessary for each measurement line.

In the literature there are various types of filters which have
been developed, for example thin film filters \cite{Vion,Courtois, Sueur},
distributed thin-film microwave filters \cite{Jin}, Thermocoax$^\circledR$ (Flers, France)
filters \cite{Zorin, Glattli} to name a few. More recently
researchers have concentrated on powder filters with increased
performance \cite{Lukashenko, Siddiqi}. Though these filters have
high performance, they are relatively bulky and cannot be used in
space constrained cryogenic systems, especially when a large
number of measurement lines are required. It was essential to
design a filter which is effective in the GHz frequency range and
that can easily be installed into space constrained cryogenic
systems. In this article we detail the fabrication of such a filter
and demonstrate its efficiency by measuring  the low temperature
superconducting transition of a granular material.

\section{Experimental Setup}
\begin{figure}[t]
\centerline{\includegraphics[height=12cm, angle = 0]{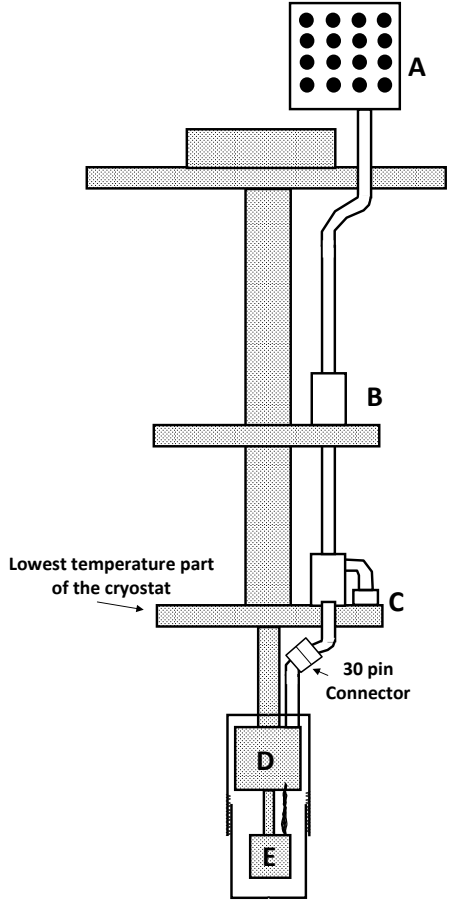}}
\caption{Schematic of our filter as installed in our cryostat. Point A in the figure shows a RF-shielded aluminium box with SMA connectors which is connected to the head of a $^3$He cryostat. Point B is the 1K plate where the wires have been thermally anchored. Point C is the base plate of the $^3$He refrigerator where the ground wires are connected. Point D and E correspond to the low temperature RC-filter and the sample holder, respectively, which form together the detachable cold finger.} \label{schematic}
\end{figure}
In figure \ref{schematic} we have shown the schematic of our filter as installed in our $^3$He cryostat.
Our filter is divided into three stages. The first stage consists
of twisted pairs of constantan wires inserted into a Cu/Ni
tube filled with ECCOSORB CRS 117 (EC) (Portion between points A and D in figure \ref{schematic}). The second stage is an
inductor embedded in the cold finger using a mixture of copper powder and
stycast (D in figure \ref{schematic}). This stage has been added to ensure efficient cooling of
the measurement lines at the lowest temperature point. In the
third stage a RC-filter is added to adjust at will the desired cut
off frequency, which, in our case is $\sim$115kHz (at the base of D in figure \ref{schematic}).

For fabricating the first stage we took thirty pairs of twisted
constantan wires, one from each pair is for the signal line and the
other is a ground line connected to the lowest temperature stage of the cryostat (point C in figure \ref{schematic}). The insulated constantan wire had a diameter of
0.12mm with a length of 1.2m. All the twisted pairs were grouped
together and pushed inside a Cu/Ni tube having an outer diameter
of 3.5mm. While putting these twisted pairs we fixed a needle-less
hypodermic syringe filled with EC at the end of the tube. This
ensured that the wires entering the tube were fully coated with
the EC. Since this process was done at a very slow rate the
material filled the tube volume completely. This tube was left
overnight for the EC to solidify and then installed into a $^3$He
refrigerator. At the room temperature end of the EC filled tube we
connected a RF shielded aluminium box with 30 SMA connectors (point A in figure \ref{schematic}). The
wires from the other end were put into a shielded copper cylindrical box
attached to the low temperature stage (1K plate) of the refrigerator for
thermal anchoring (point C in figure \ref{schematic}).

From this shielded box two sets of 30 wires went out, one from each pair in
each set. One of the two sets were grounded at the low temperature
stage in order to keep the same thermal history while the other was put into another similar but smaller
(15cm long) tube filled with EC. At the end of this small tube we
connected a home made 30 pin male connector in order to be able to
detach the cold finger containing the combined second and third
stage as shown in figure \ref{schematic}. The connector was mounted inside a copper cylinder for
shielding from environmental RF noise.

The second stage consists of 30 copper wires bundled together and
put inside a EC filled CuNi tube, which ends on the top part of the
copper cold finger (D in figure \ref{schematic}). Inside the copper piece the bundled group of
wires were loosely wound to form a coil. The total length of the
wires in the coil is 80cm. This coil was then submerged into a 1:1
mixture of white stycast (Emerson and Cuming STYCAST 1266) and copper powder which was
left to solidify overnight. The wires coming out of the coil were
soldered individually to the input of RC-filters which form the
third stage of the filter. This stage is connected to the base of point D in figure \ref{schematic}. For our measurements we have chosen the
resistance in the circuit as 511$\Omega$ and the capacitance as
2.7nF, these values can however be adjusted at will. The resistance used was MULTICOMP - MC 0.063W 0603 1\% 511R and the capacitor was muRata GRM40 SL 272J 50. These components have been tested prior to installation in order to ensure that the R/C values do not vary at low temperatures. The whole second and third stage was encased inside a RF-shielded copper tube as shown in figure.

\section{Results and Discussions}
\begin{figure}[t]
\centerline{\includegraphics[height=12cm, angle = -90]{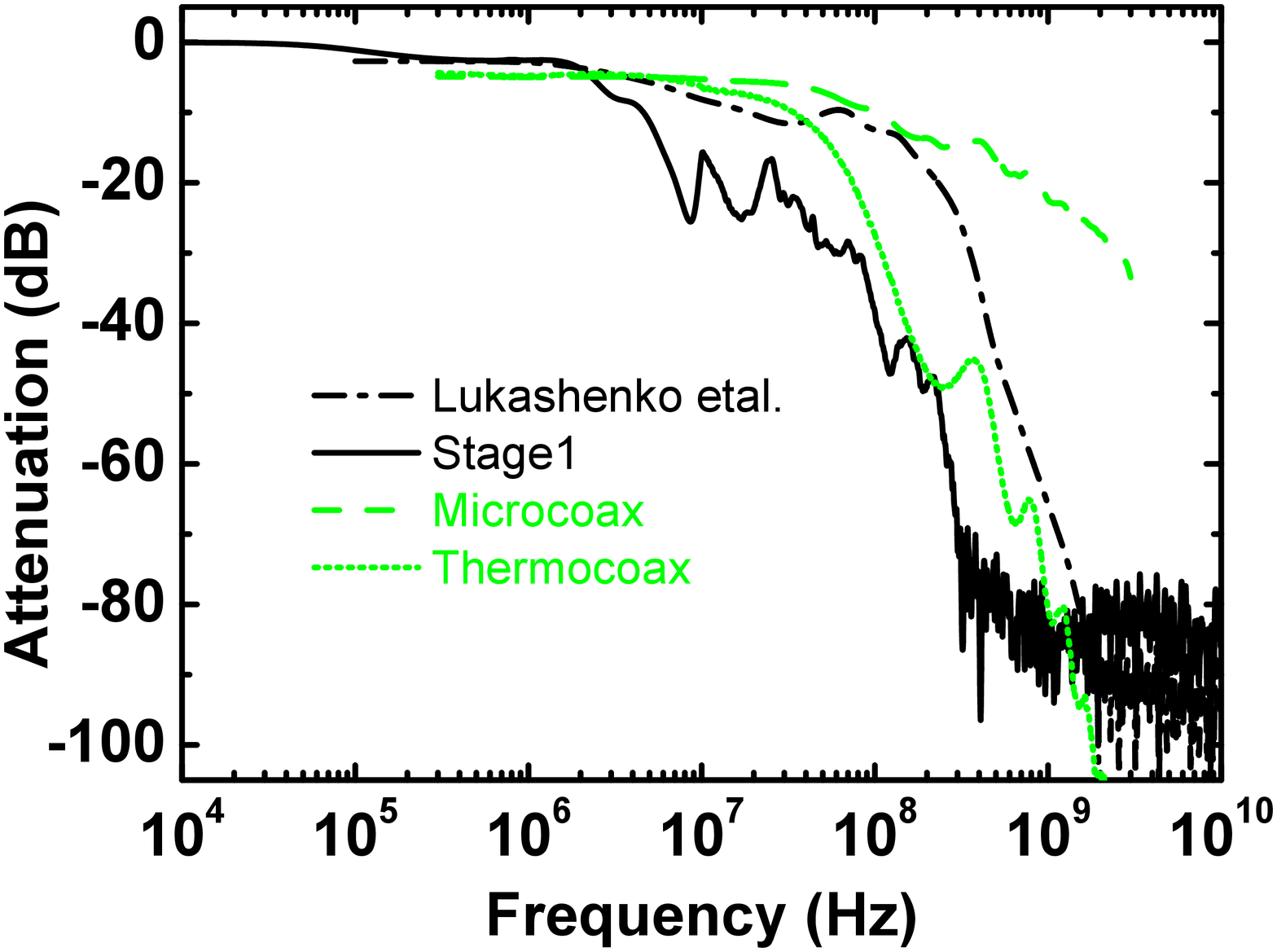}}
\caption{(Color online) A comparison between attenuation curves of
the first stage of our filter (solid black) and the filter designed
by Lukashenko et al.\cite{Lukashenko}(dash dotted black). We have also
shown the attenuation curves for thermocoax (dotted green) and
microcoax (dashed green) cables of similar lengths for
comparison.} \label{compare}
\end{figure}
\begin{figure}[b]
\centerline{\includegraphics[width=12cm, angle = 0]{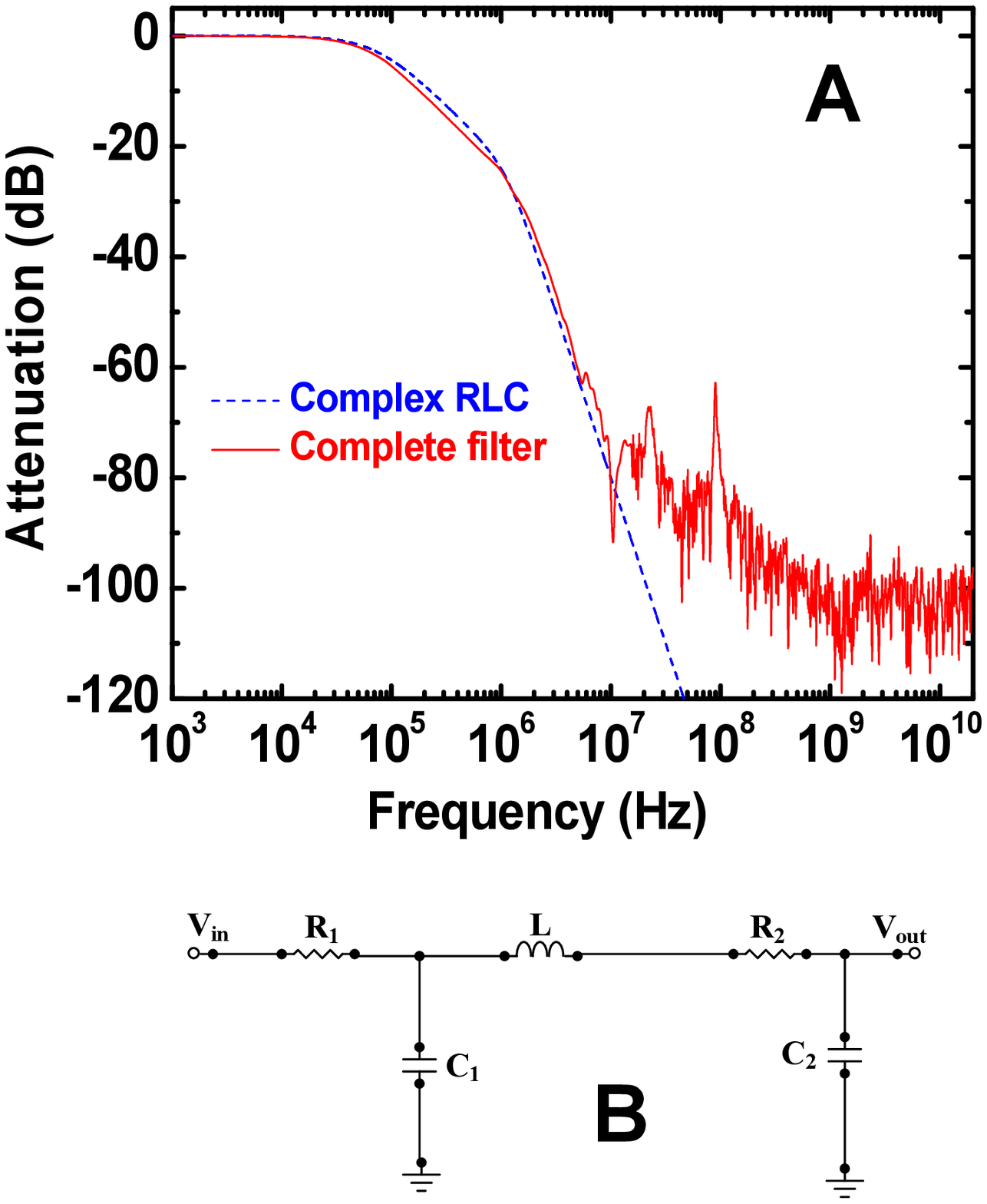}}
\caption{(Color online) Panel A shows the attenuation curve for
the complete filtered line (red solid curve). In panel B we have shown a
representative circuit which explains the obtained data at low
frequencies. The attenuation curves for this circuit is shown in
Panel A (blue dashed curve).} \label{full} \end{figure}

\begin{figure}[t]
\centerline{\includegraphics[width=8cm, angle = 0]{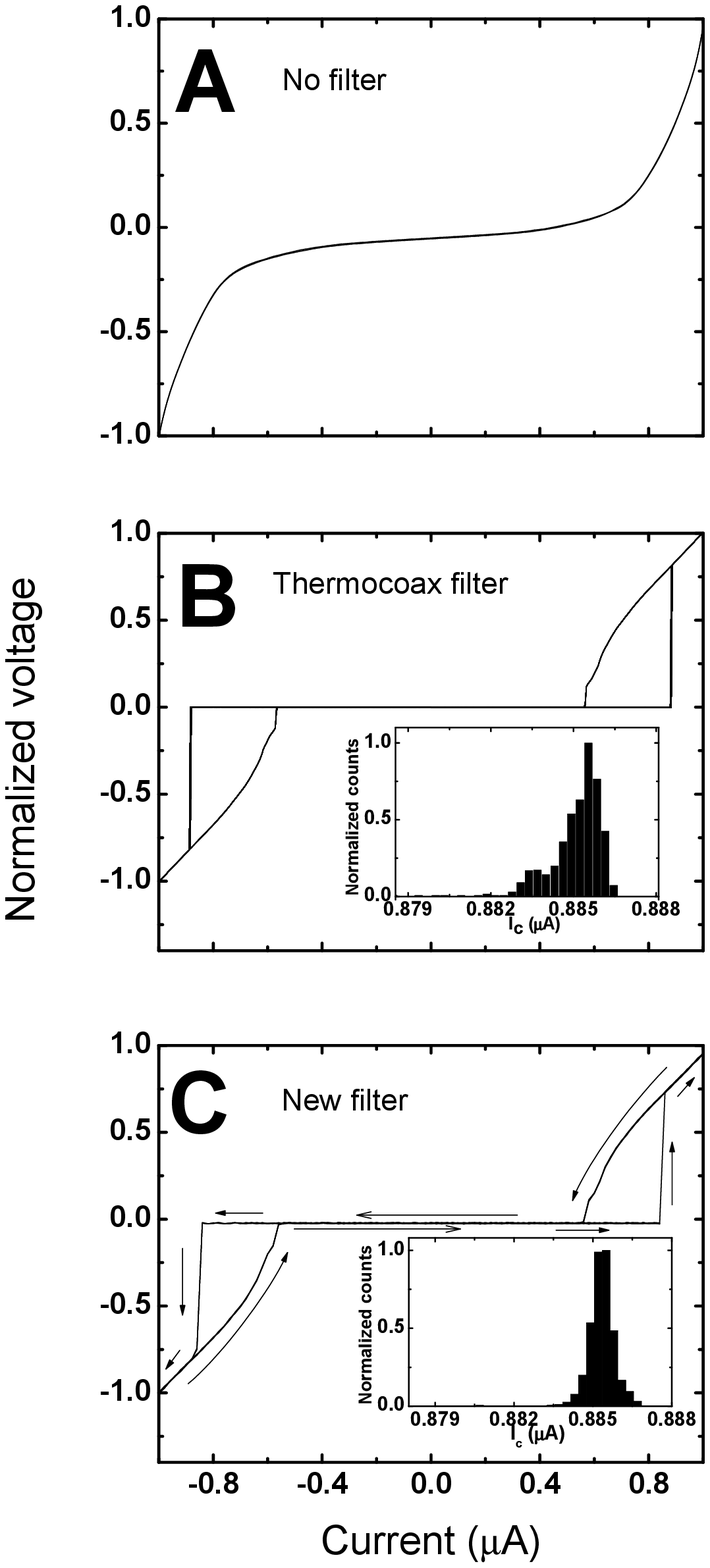}}
\caption{Voltage - current characteristic of a nanostructured diamond superconducting
device measured with and without filtered signal lines. Panel B
shows the results for lines filtered with thermocoax. The
inset shows the normalized histogram of I$_c$. Panel C shows the data measured with our filter
and the inset is the histogram of I$_c$.}
\label{iv}
\end{figure}

In figure \ref{compare} we have shown a comparison between the
attenuation of the first stage in our filter with the powder filters
of Lukashenko et al. \cite{Lukashenko}, Thermocoax and Microcoax$\circledR$ (Pottstown, PA). 
For our filters the attenuation was measured using a combination
of instruments like vector network analyzer and lock-in
amplifiers. For the high frequency part we used Agilent E8362C
vector network analyzer. The low frequency part (f $<$ 100MHz) was
recorded using a signal generator (Rohde and Schwarz SMG
Synthesized Signal Generator) and a lock-in amplifier (Signal
Recovery 7280 DSP and Stanford Research 830). For Thermo and Micro
coax cables the measurement was done using two vector network
analyzers (HP87536 for low frequency and Agilent E8362C for high
frequency) covering the whole frequency range. All three filters
had similar lengths of conducting wire of approximately 1.2 m. The
first stage in our filter had a cut off frequency of $\sim$2.1MHz
while the filters of ref. \cite{Lukashenko} had a cut-off
frequency of $\sim$4.4MHz. Microcoax had a cut off frequency of
$\sim$60MHz and the Thermocoax had a cutoff frequency of
$\sim$20MHz. One can clearly see that the performance of our first
stage is already better than any of the available options. Apart
from that, the space required to install our filter is much less
when compared with coaxial lines or the filters of ref.
\cite{Lukashenko,Kobayashi}. Though the performance of our first
stage is better than the available options, it is not good enough
for measurements involving extremely sensitive systems like the
superconducting transition of granular superconductors. We
therefore added to the first stage an inductive second stage and a
RC filter in the form of a third stage. The combined second and
third stage acts as a low pass filter whose cut-off frequency can
be adjusted.

In figure \ref{full} we have shown the attenuation curve for the
complete filter at room temperature. The line has its -3db point
at $\sim$65kHz. The attenuation at low frequencies (f $<$ 10MHz)
can be explained by a combination of complex RLC low pass filter
as indicated in the lower panel of figure \ref{full}. Resistance
R$_1$ represents the resistance of the twisted wires of the first
stage, the capacitance C$_1$ is assumed to exist between the wires
and the grounded wall of the tube containing the twisted wires,
the inductor L is from the second stage, resistance R$_2$ and
capacitance C$_2$ forms the RC-filter of the third stage. The
inductance of the coil as estimated from its geometry is
$\sim10\mu$H. R$_1$ is measured to be equal to $\sim$80$\Omega$
and R$_2$ and C$_2$ are fixed 511$\Omega$ and 2.7nF, respectively.
For this given circuit the output voltage across the capacitor
C$_2$ in terms of input voltage can be written as

\begin{eqnarray}
V_{out} &=& \left(\frac{1}{C^2+D^2}\right)^{\frac{1}{2}} V_{in} \\
\mbox{where } C &=& 1 - \omega^2LC_2-R_1R_2\omega^2C_1C_2\nonumber
\\ D &=&
R_1{\omega}C_2+R_1{\omega}C_1-R_1\omega^3LC_1C_2+R_2\omega{C_2}
\nonumber\end{eqnarray}

The simulated result for this equation is shown in Panel A (blue dashed
curve). The value of capacitor C$_1$ is determined by the best fit
curve and yields 3$\pm$1nF.

To test our system in real experiments we measured the current
voltage characteristics of a nanostructured superconducting
nanocrystalline diamond device\cite{Mandal} at 300mK  as shown in
figure \ref{iv}. Panel A represents a measurement where only
twisted wires have been employed for the electrical connection and
without any kind of filtering. Panel B shows measurements for the
lines filtered with Thermocoax and commercial filters
(Mini-circuits VLFX-80 and SLP-1.9+) while panel C shows the same
measurement using our RF filter. For the measurements, we start
sweeping the current from maximal negative current passing through
zero to maximal positive current and then sweeping the current
back to maximal negative current. Arrows in panel C indicate the
path taken by the measured voltage in the device. This process was
repeated several times generating $\sim$ 5000 cycles in all
cases. One can clearly see that while the device was not
completely superconducting due to incoming noise while measuring
with unfiltered lines, it showed the device to be completely
superconducting when being measured with filtered lines. To test
our filters in comparison with commonly used Thermocoax filters we
have also plotted the histogram of measurements of the switching current as an inset to
panels B and C. The full width at half maximum in case of
Thermocoax filtered lines is $\sim$1.1nA while for measurements
involving our filters it is $\sim$0.7nA. For the measurements using
our filters we have not used any of the commercially available filters
which have been added to increase the noise filtering of
Thermocoax lines. This clearly demonstrates that our filter is
much more effective in filtering noise detrimental to measurements
involving very small signals.

\section{Conclusion}
In conclusion we have designed a filtering system, with large
number of signal lines, which can be
installed in a space constrained cryogenic system. Our filter has performance parameters at
par with some of the best filters available in literature. We have
also given a plausible model explaining the filtering action in
our system at low frequencies (up to 10MHz). Lastly, our filters
have been tested with real experiments where we see that for the measurement of very
small critical current (below 1$\mu$A) the switching
current resolution is below 0.1\%.

\section*{Acknowledgement}
We would like to acknowledge technical assistance of Pierre Perrier as well as Clemens Winkelmann who brought our attention to ECCOSORB CRS 117(EC). This work has been partially supported by the French National Agency (ANR) in the frame of its program in `Nanosciences and Nanotechnologies' (SUPERNEMS Project no ANR-08-NANO-0 33)


\begin{thebibliography}{}
\bibitem{Vion}
D. Vion, P. F. Orfila, P. Joyez, D. Esteve and M. H. Devoret, J.
Appl. Phys. {\bf 77}, 2519 (1995)

\bibitem{Courtois}
H. Courtois, O. Buisson, J. Chaussy and B. Pannetier, Rev. Sci. Instrum. {\bf 66}, 3465 (1996)

\bibitem{Sueur}
H. Le Sueur and P. Joyez, Rev. Sci. Instrum. {\bf 77}, 115102
(2006)

\bibitem{Jin}
I. Jin, A. Amar and F. C. Wellstood, Appl. Phys. Lett. {\bf 70},
2186 (1997)

\bibitem{Zorin}
A. V. Zorin, Rev. Sci. Instrum. {\bf 66}, 4296 (1995)

\bibitem{Glattli}
D. C. Glattli, P. Jacques. A. Kumar, P. Pari and L. Saminadayar,
J. Appl. Phys. {\bf 81}, 7350 (1995)

\bibitem{Lukashenko}
A. Lukashenko and A. V. Ustinov, Rev. Sci. Instrum. {\bf 79}, 014701
(2008)

\bibitem{Siddiqi}
D. H. Slichter, O. Naaman and I. Siddiqi, Appl. Phys. Lett. {\bf
94}, 192508 (2009)

\bibitem{Kobayashi}
Masayuki Hashisaka, Yoshiaki Yamauchi, Kensaku Chida, Shuji
Nakamura, Kensuke Kobayashi and Teruo Ono, Rev. Sci. Instr. {\bf
80}, 096105 (2009)

\bibitem{Mandal}
S. Mandal, C. Naud, O. A. Williams, \'E. Bustarret, F. Omn\`es, P. Rodi\`ere, T. Meunier, L. Saminadayar and C. B\"auerle, Nanotechnology {\bf 21}, 195303 (2010)
\end{thebibliography}
\end{document}